# A COVARIATE ADJUSTMENT FOR ZERO-TRUNCATED APPROACHES TO ESTIMATING THE SIZE OF HIDDEN AND ELUSIVE POPULATIONS

By Dankmar Böhning and Peter G. M. van der Heijden

*University of Reading and Utrecht University*

In this paper we consider the estimation of population size from one-source capture–recapture data, that is, a list in which individuals can potentially be found repeatedly and where the question is how many individuals are missed by the list. As a typical example, we provide data from a drug user study in Bangkok from 2001 where the list consists of drug users who repeatedly contact treatment institutions. Drug users with 1, 2, 3, … contacts occur, but drug users with zero contacts are not present, requiring the size of this group to be estimated. Statistically, these data can be considered as stemming from a zero-truncated count distribution. We revisit an estimator for the population size suggested by Zelterman that is known to be robust under potential unobserved heterogeneity. We demonstrate that the Zelterman estimator can be viewed as a maximum likelihood estimator for a locally truncated Poisson likelihood which is equivalent to a binomial likelihood. This result allows the extension of the Zelterman estimator by means of logistic regression to include observed heterogeneity in the form of covariates. We also review an estimator proposed by Chao and explain why we are not able to obtain similar results for this estimator. The Zelterman estimator is applied in two case studies, the first a drug user study from Bangkok, the second an illegal immigrant study in the Netherlands. Our results suggest the new estimator should be used, in particular, if substantial unobserved heterogeneity is present.

**1. Introduction.** Registration files can be used to generate a list of individuals from some population of interest. If each time that an observation of a population member occurs is registered but, for one reason or another, some population members are not observed at all, the list will be incomplete and will show only part of the population. In this paper we will further de-velop a method proposed by Zelterman ([1988]) for estimating the size of a population using an incomplete list.

Consider a population of size $N$ and a count variable $Y$ taking values in the set of integers $\{0, 1, 2, 3, \ldots\}$. For example, in drug user studies $Y$ might represent the number of contacts a drug user has with the treatment institutions. Also denote with $f_0, f_1, f_2, \ldots$ the frequency with which a $0, 1, 2, \ldots$ occurs in this population. Consider now a registration where every contact with a treatment institution is registered and assume that a list of drug users is derived from this registration. Since a drug user will only be observed if there has been a positive number of contacts with the treatment institution, $y = 0$ will not be observed in the list. Hence, the list reflects a count variable truncated at zero that we denote by $Y_+$. Accordingly, the list has observed frequencies $f_1, f_2, \ldots$, but the frequency $f_0$ of zeros in the population is unknown. The size of the list is not $N$ but $n$, where $N = n + f_0$.

The distribution of the untruncated and truncated counts are connected via $P(Y_+ = j) = P(Y = j)/\{1 - P(Y = 0)\}$ for $j = 1, 2, \ldots$. For example, if $Y$ follows a Poisson distribution with parameter $\lambda$ so that

$$(1.1) \qquad P(Y = j) = Po(j \mid \lambda) = e^{-\lambda} \lambda^j / j!,$$

for $j = 0, 1, 2, \ldots$, then the associated distribution of $Y_+$ is given as

$$(1.2) \qquad P(Y_+ = j) = Po_+(j \mid \lambda) = \frac{e^{-\lambda}}{1 - e^{-\lambda}} \lambda^j / j!,$$

with $j = 1, 2, 3, \ldots$.

Given that all units of the population have the same probability $P_i(Y > 0) = P(Y > 0) = 1 - P(Y = 0)$ of being included in the list, the population size can be estimated by means of the Horvitz–Thompson estimator

$$(1.3) \qquad \hat{N} = \sum_{i=1}^{n} \frac{1}{P_i(Y > 0)} = \frac{n}{1 - P(Y = 0)} = \frac{n}{1 - g(\lambda)},$$

where $g(\lambda) = e^{-\lambda}$, or more generally, $g(\lambda)$ is the probability of a zero count for a given count distribution. For more details on this type of capture–recapture methodology, see van der Heijden et al. ([2003a]), van der Heijden, Cruyff and van Houwelingen ([2003b]), Böhning and Schön ([2005]), Roberts and Brewer ([2006]) or McKendrick ([1926]) (for a historic account). General introductions to capture–recapture are found in Bishop, Fienberg and Holland ([1975]), Hook and Regal ([1995]) and the contributions of the International Working Group for Disease Monitoring and Forecasting ([1995a], [1995b]).

In what follows we further develop an estimator for $\lambda$ proposed by Zelterman ([1988]), which can be used in ([1.3]) to obtain a population size estimate. This estimator for $\lambda$ uses limited information from the observed count distribution to arrive at an estimate of the population size, making it robust. Our key extension to this estimator for $\lambda$ is to put it into a maximum likelihood framework, which allows further development using a regression framework. In Section 2 we review the Zelterman estimator, including its robustness properties. In Section 3 we demonstrate that the Zelterman estimator is a maximum likelihood estimator and use this result to estimate its variance and generalize the estimator to accommodate covariates. Section 4 points out the connections to Chao's estimator. The paper concludes with a case study section where we utilize examples from a Bangkok illicit drug user study and a reanalysis of illegal immigrant data analyzed earlier by van der Heijden et al. (2003a).

**2. The Zelterman estimator.** In equation (1.3) we used the Horvitz–Thompson approach to arrive at an estimate of the population size. This approach requires that $\lambda$ is known and if it is not, it needs to be estimated. Clearly, $\lambda$ can be estimated with maximum likelihood under the assumption of a homogeneous truncated Poisson distribution. Instead of estimating $\lambda$ under the assumption of a homogeneous Poisson distribution, Zelterman (1988) argued that the Poisson assumption might not be valid over the entire range of possible values for $Y$ but it might be valid for small ranges of $Y$ such as from $j$ to $j+1$, so that it would be meaningful to use only the frequencies $f_j$ and $f_{j+1}$ in estimating $\lambda$. Since for any $j$ both the truncated as well as the untruncated Poisson distribution have the property that $Po(j+1 \mid \lambda)/Po(j \mid \lambda) = \lambda/(j+1)$ and $Po_+(j+1 \mid \lambda)/Po_+(j \mid \lambda) = \lambda/(j+1)$, respectively [see equations (1.1) and (1.2)], $\lambda$ can be written as

$$(2.1) \qquad \lambda = \frac{(j+1)Po(j+1 \mid \lambda)}{Po(j \mid \lambda)} = \frac{(j+1)Po_+(j+1 \mid \lambda)}{Po_+(j \mid \lambda)}.$$

An estimator for $\lambda$ is obtained by replacing $Po_+(j \mid \lambda)$ by the empirical frequency $f_j$:

$$(2.2) \qquad \hat{\lambda}_j = \frac{(j+1)f_{j+1}}{f_j}.$$

If $j = 1$, we find $\hat{\lambda}_1 = 2f_2/f_1$, and this estimator is often considered for two reasons: for one, $\hat{\lambda}_1$ is using frequencies in the vicinity of $f_0$ which is the target of prediction, and two, in many application studies for estimating $f_0$ the majority of counts fall into $f_1$ and $f_2$. Clearly, the estimator is *unaffected* by changes in the data for counts larger than 2, which contributes largely to its robustness. We will call $\hat{\lambda}_1 = 2f_2/f_1$ the *Zelterman estimator for* $\lambda$ and, when this estimate is used in (1.3), this leads to the *Zelterman estimator of the population size,* $\hat{N}$. If the context is clear, we will simply use the term *Zelterman estimator*.

The Zelterman estimator is an estimator which is very simple to understand and to use and this might be one of the reasons why it is quite popular in applications such as drug user studies [Hay and Smit (2003), Van Hest et al. (2007)]. It is also thought of as being less sensitive to model violations than the estimator that is derived under the assumption of the homogeneous Poisson distribution, that uses the entire range of frequencies $f_j$. Indeed, the Zelterman estimator also works rather well with contaminated distributions as given by mixtures or approximated by mixtures [compare Zelterman (1988)]. We now look at a study to illustrate the application of the estimator.

TABLE 1
*Frequency distribution $f_y$ of Metamphetamine users with exactly y repeated contacts with treatment institutions*

| $y$ | 1 | 2 | 3 | 4 | 5 | 6 | 7 | 8 | 9 | 10 | 11 | 12 |
|-----|---|---|---|---|---|---|---|---|---|----|----|----|
| $f_y$ | 3114 | 163 | 23 | 20 | 9 | 3 | 3 | 3 | 4 | 3 | 0 | 1 |

EXAMPLE: (*Estimating the number of Metamphetamine-users in Bangkok*, 2001). Let us illustrate the estimator for a data set of users of Metamphetamine in Bangkok [Böhning et al. (2004)]. The distribution of contact counts with treatment institutions is provided in Table 1.

In total 3346 users were observed. We find $\hat{\lambda}_1 = (2 \times 163)/3114 = 0.1047$ (with 95% CI 0.0894–0.1225) and, using (1.3), this gives $\hat{N} = 3346/(1 - \exp(0.1047)) = 33,664$ (CI 28,520–38,808). The observed/hidden ratio equals $3346/(33,664-3346) = 0.1104$ and the completeness is $3346/33,664 = 0.0994$. Note that the maximum likelihood estimator derived under the homogeneous Poisson assumption is $\hat{\lambda} = 0.2463$ (CI 0.2245–0.2703), leading to a population size estimate of $\hat{N} = 3346/(1 - \exp(-0.2463)) = 15,325$ (CI 13,989–16,661), which differs considerably from the Zelterman estimator of the population size. The confidence intervals are based upon normal approximations using a variance expression given in Section 3.1 below. Since it is reasonable to assume that the counts stem from a contaminated distribution rather than from a homogeneous distribution, the Zelterman estimate appears to be more reasonable. In addition, the homogeneous Poisson estimate is biased downward if heterogeneity is present [van der Heijden et al. (2003a), van der Heijden, Cruyff and van Houwelingen (2003b), Böhning and Schön (2005)], so that a strong disagreement of the homogeneous Poisson estimate to the Zelterman estimate might be taken as an indication for a lack of fit for the homogeneous Poisson as occurs here. In such cases, the Zelterman estimate will be the better choice.

**3. The Zelterman estimator as a maximum likelihood estimator.** In this section we will show that the Zelterman estimator is also a maximum likelihood estimator. It is based upon the observation that a Poisson distribution with parameter $\lambda$ constrained to values $Y = 1$ and $Y = 2$ yields a binomial distribution with parameter $p = (\lambda/2)/(1 + \lambda/2) = \lambda/(2 + \lambda)$. This result will allow for a simple derivation of the variance (see Section 3.1), as well as an extension of the Zelterman estimator that allows for covariates (see Section 3.2).

3.1. *A likelihood for the Zelterman estimator.* If we consider the probability for a count of 1 and a count of 2 given as $e^{-\lambda}\lambda/(e^{-\lambda}\lambda + e^{-\lambda}\lambda^2/2)$ and $(e^{-\lambda}\lambda^2/2)/(e^{-\lambda}\lambda + e^{-\lambda}\lambda^2/2)$, respectively, we see that after some simplification we have the likelihood

$$(3.1) \qquad \left(\frac{2}{2+\lambda}\right)^{f_1} \times \left(\frac{\lambda}{2+\lambda}\right)^{f_2} = (1-p)^{f_1} p^{f_2},$$

which is proportional to a *binomial likelihood* with event parameter $p = \lambda/(2 + \lambda)$, the probability for $Y = 2$. This binomial likelihood is maximized for $\hat{p} = f_2/(f_1 + f_2)$. In addition, as $\lambda$ is connected uniquely to $p$ via $\lambda = 2p/(1 - p)$, the invariance property of maximum likelihood estimators yields $\hat{\lambda}_1 = 2f_2/f_1$ as a maximum likelihood estimate with respect to the likelihood (3.1). We summarize this in the following theorem.

THEOREM 1. *Consider a Poisson count $Y$ where all counts are truncated unless $Y = 1$ or $Y = 2$. Then:*

(a) *the associated likelihood is given by (3.1),*
(b) *the maximum likelihood estimator with respect to (3.1) is*

$$\hat{p} = f_2/(f_1 + f_2) \quad or \quad \hat{\lambda}_1 = 2f_2/f_1.$$

One of the first benefits of identifying the Zelterman estimator $\hat{\lambda}_1$ as a truncated maximum likelihood estimator is the fact that its variance is readily available as $\text{Var}(\hat{p}) = p(1 - p)/(f_1 + f_2)$, which can be estimated as $f_2 f_1/(f_1 + f_2)^3$. Now, $\hat{\lambda}_1 = 2\frac{\hat{p}}{1-\hat{p}}$, and using a first order $\delta$-method,

$$\text{Var}\log(\hat{\lambda}_1) = \text{Var}(\log\hat{p} - \log(1-\hat{p})) \approx \left(\frac{1}{p} + \frac{1}{1-p}\right)^2 \frac{p(1-p)}{f_1 + f_2}$$

and, finally, plugging in an estimate for $p$, we arrive at

$$(3.2) \qquad \widehat{\text{Var}}\log(\hat{\lambda}_1) \approx \left(\frac{f_2}{f_1 + f_2}\frac{f_1}{f_1 + f_2}(f_1 + f_2)\right)^{-1} = \frac{1}{f_1} + \frac{1}{f_2},$$

leading to a simple closed form expression for the variance of the logarithm of the Zelterman estimator. In addition, using a first order $\delta$-method, we have that $\text{Var}\log\hat{\lambda}_1 \approx \frac{1}{\lambda^2}\text{Var}\,\hat{\lambda}_1$, which can be rephrased as

$$\text{Var}\,\hat{\lambda}_1 \approx \lambda^2 \,\text{Var}\log\hat{\lambda}_1.$$

Plugging in the Zelterman estimate for $\lambda$ leads to the result (b) in the following theorem.

THEOREM 2.    *Consider a situation as in Theorem 1. Then:*

(a) $\widehat{\mathrm{Var}} \log(\hat{\lambda}_1) \approx \frac{1}{f_1} + \frac{1}{f_2}$,

(b) $\widehat{\mathrm{Var}}(\hat{\lambda}_1) \approx \frac{4f_2(f_1+f_2)}{f_1^3}$.

3.2. *Extension of the likelihood for covariates.*    A second benefit of identifying the Zelterman estimator as a truncated maximum likelihood estimator is that it is now easy to incorporate covariates into the modeling process. Let $Z$ be a binary indicator variable indicating $Z = 1$ if $Y = 2$ and $Z = 0$ if $Y = 1$. Then, the likelihood (3.1) can be written as

$$(3.3) \quad \prod_{i=1}^{f_1+f_2} p_i^{z_i}(1-p_i)^{1-z_i} = \prod_{i=1}^{f_1+f_2} \left( \frac{\lambda_i/2}{1+\lambda_i/2} \right)^{z_i} \left( 1 - \frac{\lambda_i/2}{1+\lambda_i/2} \right)^{1-z_i}.$$

Suppose that covariates are available in the form of a vector $\mathbf{x}_i$ for the $i$th unit in the list. In a generalized linear model (logistic regression model) connecting the binary outcome probability $p_i$ with the *linear predictor* $\eta_i = \beta^T \mathbf{x}_i$ with a logit link, we have that

$$p_i = \frac{e^{\eta_i}}{1+e^{\eta_i}}.$$

On the other hand, $p_i$ and the local Poisson parameter $\lambda_i$ are connected via

$$p_i = \frac{\lambda_i/2}{1+\lambda_i/2},$$

so that $\lambda_i$ and the linear predictor $\eta_i$ are simply connected via $\lambda_i/2 = e^{\eta_i}$ or $\lambda_i = 2e^{\eta_i}$. Note that the binary response probability $P(Z_i = 1) = p_i$ is connected to the linear predictor $\eta_i$ via the *logistic link* function, whereas the Poisson mean $\lambda_i = 2e^{\eta_i}$ uses the *log link* function, that is, both are generalized linear models using the canonical link functions.

Maximum likelihood estimation can use existing tools for logistic regression. All that is needed is to regress the binary outcomes $z_1, \ldots, z_n$ on $\mathbf{x}_i$ to find the MLE $\hat{\beta}$ of $\beta$. This provides the predicted probabilities $\hat{p}_i = e^{\hat{\eta}_i}/(1+e^{\hat{\eta}_i})$ and the Zelterman estimates of parameters $\lambda_i$ are obtained as $2\hat{p}_i/(1-\hat{p}_i)$.

In order to derive a generalized Zelterman estimator of the population size $N$ under this framework, we can employ the Horvitz–Thompson approach in the following way:

$$\hat{N}_Z = \sum_{i=1}^{n} \frac{1}{1-\exp(-\hat{\lambda}_i)}$$

(3.4)
$$= \sum_{i=1}^{n} \frac{1}{1 - \exp(-2\hat{p}_i/(1 - \hat{p}_i))} = \sum_{i=1}^{n} \frac{1}{1 - \exp(-2e^{\hat{\eta}_i})}.$$

In addition, it is possible to find an estimate of the variance of the generalized Zelterman estimator (3.4) which we write as

$$\hat{N}_Z = \sum_{i=1}^{n} \frac{1}{w_i} = \sum_{i=1}^{N} \frac{\Delta_i}{w_i},$$

where $w_i = 1 - \exp(-2e^{\hat{\eta}_i})$ and $\Delta_i$ is an indicator which is 1 (unit is sampled) with probability $w_i$ and 0 (unit is not sampled) with probability $1 - w_i$. Note that $w_i = 1 - \exp(-2e^{\hat{\eta}_i})$ is not fixed, but a random quantity itself. This excludes the direct application of known variance formulas for the Horvitz–Thompson estimator and their variations such as Sen–Yates–Grundy [for details, see Thompson (2002), pages 54–55]. Variance estimation of the Horvitz–Thompson estimator with estimated $w_i$ (which might no longer be called the Horvitz–Thompson estimator) needs to take into account the variability in estimating the linear predictor $\hat{\eta}_i$. This problem was first pointed out by Huggins (1989). To accomplish the task, we use the techniques of *conditional moments* [see Ross (1985), page 125] and results from van der Heijden et al. (2003a). Details are in the Appendix. We state here only the final variance approximation:

(3.5)  $$\widehat{\mathrm{Var}}(\hat{N}_Z) \approx \sum_{i=1}^{n} (1 - w_i)/w_i^2 + \sum_{i=1}^{n} \left( \frac{(1 - w_i)v_i}{w_i^2} \right)^2 \mathbf{x}_i^T \, Cov(\hat{\beta}) \mathbf{x}_i,$$

where $w_i = 1 - \exp(v_i)$ and $v_i = -2e^{\hat{\eta}_i}$, so that $w_i = w_i(\hat{\beta}) = 1 - \exp(v_i) = 1 - \exp(-2e^{\hat{\eta}_i}) = 1 - \exp(-2e^{\hat{\beta}^T \mathbf{x}_i})$.

## 4. The connection to Chao's estimator.

In this section we point out some connections to another population size estimator proposed by Chao (1987, 1989) that also uses only the counts $f_1$ and $f_2$. We provide these results because generalizing this estimator into a maximum likelihood framework was less successful. Chao suggested the estimator $\hat{N}_C = n + \frac{f_1^2}{2f_2}$. The estimator is based upon the Cauchy–Schwarz inequality [see also Wilson and Collins (1992)] for the nonparametric mixture of a Poisson, namely,

$$\left( \int_0^\infty \lambda e^{-\lambda} \, d\lambda \right)^2 \leq \int_0^\infty e^{-\lambda} \, d\lambda \int_0^\infty \lambda^2 e^{-\lambda} \, d\lambda,$$

where the inequality of the Cauchy–Schwarz $(\int uv)^2 \leq (\int u^2)(\int v^2)$ is used with $u(\lambda) = \sqrt{e^{-\lambda}}$ and $v(\lambda) = \lambda\sqrt{e^{-\lambda}}$ and leading to $p_1^2 \leq p_0 \times 2p_2$, so that $\frac{f_1^2}{2f_2}$ estimates a lower bound for $f_0$. Chao (1987, 1989) suggests to use this bound as an estimator if higher frequency counts are small. It is mentioned
frequently in the applied statistical literature [see, e.g., Smit, Reinking and
Reijerse (2002)] that the Zelterman estimator and Chao's estimator are often
quite close. Indeed, if we compute the Chao estimator in our drug user
example, we find that $\hat{N}_C = 3346 + 3114^2/(2 \times 163) = 33{,}091$ (95 percent
CI is 28,058–38,124), which is not far from $\hat{N}_Z = 33{,}664$. Furthermore, it is
often claimed that the Zelterman estimator is usually larger than Chao's
estimator as it is in our example here. Hence, it is interesting to investigate
the relationship between the two estimators more theoretically.

The Zelterman estimator and the Chao estimator are connected as follows.
Let us write the Zelterman estimate for $f_0$ as

$$n \frac{\exp(-\hat{\lambda})}{1 - \exp(-\hat{\lambda})} = \frac{n}{\exp(\hat{\lambda}) - 1} \approx \frac{n}{\hat{\lambda} + 1/2\hat{\lambda}^2},$$

using the first three terms of the MacLaurin series for the exponential func-
tion: $\exp(x) = 1 + x + \frac{1}{2}x^2 + \cdots$. This can be further written as

$$\frac{n}{\hat{\lambda} + 1/2\hat{\lambda}^2} = \frac{f_1^2}{2f_2} \frac{n}{f_1 + f_2} \geq \frac{f_1^2}{2f_2},$$

the latter being Chao's lower bound estimate of $f_0$. Two statements follow
now easily from this representation and are summarized in Theorem 3 below.

THEOREM 3. *Consider a situation as in Theorem 1:*

(a) *Assume that* $\frac{n}{f_1 + f_2} > 1$. *Then, for any* $\varepsilon > 0$ *exists* $\delta > 0$ *such that*

$$if \ \frac{f_2}{f_1} < \delta, \ then \qquad \hat{N}_C \leq \hat{N}_Z + \varepsilon.$$

(b) *If* $\frac{n}{f_1 + f_2} = 1$, *then* $\hat{N}_Z \geq \hat{N}_C$ *and* $\hat{N}_Z - \hat{N}_C = \mathcal{O}(\hat{\lambda}_1^3)$.

Zelterman's estimator is not always larger than Chao's. Note that state-
ment (b) gives a condition which leads to Chao's estimator being larger than
the one of Zelterman. Statement (b) follows from the fact that $n/[\exp(\hat{\lambda}) -
1] \leq n/(\hat{\lambda} + \frac{1}{2}\hat{\lambda}^2)$ for any nonnegative $\hat{\lambda}$. The term $n/(\hat{\lambda} + \frac{1}{2}\hat{\lambda}^2)$ simplifies to
$f_1^2/(2f_2)[n/(f_1 + f_2)] = f_1^2/(2f_2)$ and the result follows. The second part of
statement (b) follows from the fact that

$$\exp(\hat{\lambda}) - 1 = \sum_{i=1}^{\infty} \hat{\lambda}^i/i! = \hat{\lambda} + \hat{\lambda}^2/2 + \lambda^3(1/3! + \hat{\lambda}/4! + \cdots),$$

where the left-hand side corresponds to the Zelterman estimator and the
first two terms of the right-hand side correspond to the Chao estimator.
This ends the proof.

Note the difference between statements (a) and (b) in Theorem 3. Statement (b) says that the estimators of Chao and Zelterman are close, with Chao's estimator larger than the one of Zelterman if the ratio of the count of twos to the count of ones is small *and* the proportion of both of them among all observations is close to one. Statement (a) says that the estimator of Chao is bounded above by the estimator of Zelterman (but they need not to be close) if the ratio of the count of twos to the count of ones is small.

Some elementary calculations show that $\hat{N}_C = n + \frac{f_1^2}{2f_2}$ also satisfies

$$(4.1) \qquad \hat{N}_C = \frac{n}{1 - \hat{p}_1^2/(2\hat{p}_2)} = \frac{n}{1 - f_1^2/(2f_2\hat{N}_C)},$$

where $\hat{p}_j = f_j/\hat{N}_C$ for $j = 1, 2$. Unfortunately, (4.1) contains $\hat{N}_C$ on both sides of the equation, which causes difficulties when we aim to generalize this for data with covariate information. More details on this aspect of Chao's estimator are available from the authors upon request.

## 5. Examples.

5.1. *The Bangkok drug users study example.* We will illustrate the generalized Zelterman approach using the Bangkok drug users study [Böhning et al. (2004)] introduced in Section 2. Let us consider the female drug users only. Tables 2 and 3 show the distribution of contact counts to treatment institutions by age for Metamphetamine and Heroin users respectively. These are very different subpopulations of the drug user population in the Bangkok metropolis, as indicated by the quite different age distributions. Clearly, the age distribution of the Metamphetamine users is younger than the age distribution of the Heroin users (see Tables 2 and 3). To analyze these data, STATA and GAUSS macros are available in the supplemental articles Böhning and van der Heijden (2008a, 2008b). The results of the analysis are provided in Table 4. None of the subpopulations seems to be affected by age as follows from a likelihood ratio test. Accordingly, the population size estimates, unadjusted and adjusted for age, do not differ much. Whereas for the female Heroin user population a completeness of identification of about 50% is reached (268/504), the completeness of identification is less than 10% for the Metamphetamine users (274/3714).

5.2. *The illegal immigrant's study.* As a second example, we discuss the estimation of the number of illegal immigrants in four large cities in the Netherlands from police records, analyzed with the truncated Poisson regression model by van der Heijden et al. (2003a). In their analysis van der Heijden et al. focus on those illegal immigrants that, once apprehended, cannot be effectively expelled by the police because, for example, their home country does not cooperate with the organization of deportation. In such cases the police request the individuals to leave the country, but it is unlikely that they will abide by such a request. Hence, they can be apprehended multiple times. The data contain four covariates: gender, age, home country and reason for being arrested (or rearrested). For details about the data we refer to van der Heijden et al. (2003a). The observed frequencies for the covariate categories can be found in Table 5 and are reproduced from van der Heijden et al. (2003a). The data are provided in a supplemental file [Böhning and van der Heijden (2008c)].

TABLE 2
*Distribution of repeated contact counts y to treatment institutions of female Metamphetamine users by age*

| Age | # users with $y$ contacts: | | | | All |
|-----|-----|-----|-----|-----|-----|
| | **1** | **2** | **3** | **4** | **All** |
| 13 | 3 | 0 | 0 | 0 | 3 |
| 14 | 5 | 0 | 0 | 0 | 5 |
| 15 | 23 | 0 | 0 | 0 | 23 |
| 16 | 18 | 1 | 0 | 0 | 19 |
| 17 | 19 | 1 | 0 | 0 | 20 |
| 18 | 21 | 1 | 1 | 0 | 23 |
| 19 | 23 | 1 | 0 | 0 | 24 |
| 20 | 23 | 0 | 0 | 0 | 23 |
| 21 | 17 | 0 | 1 | 0 | 18 |
| 22 | 22 | 1 | 0 | 0 | 23 |
| 23 | 10 | 2 | 0 | 0 | 12 |
| 24 | 15 | 0 | 0 | 0 | 15 |
| 25 | 13 | 2 | 0 | 0 | 15 |
| 26 | 12 | 0 | 0 | 0 | 12 |
| 27 | 6 | 0 | 0 | 0 | 6 |
| 28 | 4 | 0 | 0 | 0 | 4 |
| 29 | 4 | 0 | 0 | 0 | 4 |
| 30 | 5 | 0 | 0 | 0 | 5 |
| 31 | 4 | 0 | 0 | 0 | 4 |
| 32 | 1 | 0 | 0 | 0 | 1 |
| 33 | 1 | 1 | 0 | 0 | 2 |
| 34 | 2 | 0 | 0 | 0 | 2 |
| 35 | 2 | 0 | 0 | 0 | 2 |
| 36 | 3 | 0 | 0 | 1 | 4 |
| 37 | 3 | 0 | 0 | 0 | 3 |
| 38 | 1 | 0 | 0 | 0 | 1 |
| 39 | 1 | 0 | 0 | 0 | 1 |
| All | 261 | 10 | 2 | 1 | 274 |

In Table 6 we provide the estimates of both the truncated Poisson regression model as well as the Zelterman regression model. Both models provide

TABLE 3

*Distribution of repeated contact counts $y$ to treatment institutions of female Heroin users by age*

| Age | # users with $y$ contacts: | | | | All |
|-----|---|---|---|---|-----|
| | **1** | **2** | **3** | **4** | |
| 16 | 0 | 1 | 0 | 0 | 1 |
| 17 | 1 | 0 | 0 | 0 | 1 |
| 18 | 3 | 0 | 0 | 1 | 4 |
| 19 | 1 | 1 | 1 | 2 | 5 |
| 20 | 0 | 3 | 2 | 2 | 7 |
| 21 | 6 | 0 | 0 | 7 | 13 |
| 22 | 3 | 5 | 1 | 5 | 14 |
| 23 | 10 | 3 | 2 | 9 | 24 |
| 24 | 11 | 4 | 4 | 9 | 28 |
| 25 | 8 | 0 | 1 | 2 | 11 |
| 26 | 13 | 4 | 3 | 4 | 24 |
| 27 | 6 | 0 | 1 | 7 | 14 |
| 28 | 4 | 1 | 2 | 3 | 10 |
| 29 | 4 | 3 | 1 | 2 | 10 |
| 30 | 0 | 2 | 1 | 2 | 5 |
| 31 | 3 | 1 | 2 | 3 | 9 |
| 32 | 4 | 1 | 0 | 1 | 6 |
| 33 | 6 | 1 | 3 | 1 | 11 |
| 34 | 2 | 2 | 0 | 3 | 7 |
| 35 | 2 | 0 | 1 | 0 | 3 |
| 36 | 2 | 1 | 0 | 3 | 6 |
| 37 | 3 | 3 | 1 | 1 | 8 |
| 38 | 3 | 1 | 1 | 2 | 7 |
| 39 | 0 | 2 | 0 | 0 | 2 |
| 40 | 4 | 2 | 1 | 0 | 7 |
| 41 | 1 | 2 | 1 | 1 | 5 |
| 42 | 4 | 0 | 0 | 1 | 5 |
| 43 | 2 | 0 | 0 | 1 | 3 |
| 44 | 2 | 0 | 2 | 1 | 5 |
| 45 | 1 | 0 | 0 | 1 | 2 |
| 46 | 0 | 0 | 0 | 1 | 1 |
| 47 | 2 | 1 | 0 | 0 | 3 |
| 48 | 1 | 0 | 1 | 0 | 2 |
| 49 | 1 | 0 | 0 | 1 | 2 |
| 52 | 1 | 0 | 0 | 0 | 1 |
| 58 | 1 | 0 | 0 | 0 | 1 |
| 62 | 1 | 0 | 0 | 0 | 1 |
| All | 116 | 44 | 32 | 76 | 268 |

TABLE 4
*Estimated population size of female drug users in Bangkok with 95% confidence interval without and with adjustment for age of drug user, and logistic log-likelihood LL*

| Drug | Covariates | $\hat{N}_Z$ (95% CI) | $LL$ |
|------|-----------|---------------------|------|
| Heroin | None | 504 (389–628) | −94.11 |
| | AGE | 505 (379–630) | −93.86 |
| Metamphetamine | None | 3714 (1417–6011) | −42.81 |
| | AGE | 3772 (1376–6169) | −42.72 |

TABLE 5
*Illegal immigrants not effectively expelled. Observed frequencies for covariate categories*

| Covariate category | $f_1$ | $f_2$ | $f_3$ | $f_4$ | $f_5$ | $f_6$ | Total |
|--------------------|-------|-------|-------|-------|-------|-------|-------|
| >40 years | 105 | 6 | | | | | 111 |
| <40 years | 1540 | 177 | 37 | 13 | 1 | 1 | 1769 |
| Female | 366 | 24 | 6 | 1 | 1 | | 398 |
| Male | 1279 | 159 | 31 | 12 | | 1 | 1482 |
| Turkey | 90 | 3 | | | | | 93 |
| North Africa | 838 | 146 | 28 | 9 | 1 | 1 | 1023 |
| Rest Africa | 229 | 11 | 3 | | | | 243 |
| Surinam | 63 | 1 | | | | | 64 |
| Asia | 272 | 9 | 1 | 2 | | | 284 |
| America, Australia | 153 | 13 | 5 | 2 | | | 173 |
| Being illegal | 224 | 29 | 5 | 1 | | | 259 |
| Other reason | 1421 | 154 | 32 | 12 | 1 | 1 | 1621 |

similar point estimates, but the estimated standard errors of the Zelterman regression model are somewhat larger than those of the truncated Poisson regression model, yielding less parameter estimates in the Zelterman regression model deviating significantly from zero.

In Table 7 we present the population size estimates for a series of models. The top panel has been reproduced from van der Heijden et al. (2003a). It shows that the truncated Poisson regression model with covariates Gender, Age and Nation provides the best fitting main effects model both in terms of deviance as well as AIC, and when these three variables are included Reason does not increase the fit significantly. The population size estimate is 12,690 with a CI of (7186–18,194).

Interestingly, the top panel provides for each model a Lagrange multiplier test [Gurmu (1991)] that can be used to test for overdispersion in the zero-truncated Poisson regression model as a result of unobserved heterogeneity. This test compares the model fit of the Poisson model with alternative models with an extra dispersion parameter included, such as the negative binomial regression model. Van der Heijden et al. (2003a) and Böhning and Schön (2005) show that, if there is evidence for unobserved heterogeneity in a model, the population size estimate will underestimate the true population size [see also Böhning and Kuhnert (2006)]. For the illegal immigrant data this appears to be the case for every model in the top panel of Table 7.

Table 6

*Truncated Poisson regression model (columns 1 and 2) and Zelterman regression model (columns 3 and 4) fit to the illegal immigrants data*

| Regression parameters | MLE | SE | MLE-Z | SE-Z |
|---|---|---|---|---|
| Intercept | −2.317 | 0.449 | −3.359 | 0.528 |
| Gender (male = 1, female = 0) | 0.397 | 0.163 | 0.535 | 0.232 |
| Age (<40 yrs = 1, >40 yrs = 0) | 0.975 | 0.408 | 0.567 | 0.434 |
| Nationality | | | | |
| (Turkey) | −1.675 | 0.603 | −1.030 | 0.657 |
| (North Africa) | 0.190 | 0.194 | 0.579 | 0.307 |
| (Rest of Africa) | −0.911 | 0.301 | 0.664 | 0.425 |
| (Surinam) | −2.337 | 1.014 | −1.720 | 1.050 |
| (Asia) | −1.092 | 0.302 | −1.056 | 0.448 |
| (America and Australia) | 0.000 | | 0.000 | |
| Reason (being illegal = 1, else = 0) | 0.011 | 0.162 | 0.189 | 0.220 |

Table 7

*Estimates $\hat{N}$ and 95% confidence intervals for N obtained from fitting different truncated Poisson regression models (first five models) and Zelterman regression models (last five models). Model comparisons using the likelihood-ratio test and AIC-criterion are also given. $\chi^2_{(1)}$ is the Lagrange multiplier test testing for overdispersion in the Poisson regression model*

| | AIC | $G^2$ | df | $P^*$ | $\chi^2_{(1)}$ | $\hat{N}$ | CI |
|---|---|---|---|---|---|---|---|
| Poisson regression | | | | | | | |
| Null | 1805.9 | | | | 106.0 | 7080 | 6363–7797 |
| G | 1798.3 | 9.6 | 1 | 0.002 | 99.7 | 7319 | 6504–8134 |
| G + A | 1789.0 | 11.2 | 1 | <0.001 | 93.7 | 7807 | 6637–8976 |
| G + A + N | 1712.9 | 86.1 | 5 | <0.001 | 55.0 | 12,690 | 7186–18,194 |
| G + A + N + R | 1714.9 | 0.004 | 1 | 0.949 | 55.0 | 12,691 | 7185–18,198 |
| Zelterman regression | | | | | | | |
| Null | 1191.4 | | | | | 9424 | 8084–10,765 |
| G | 1184.3 | 9.1 | 1 | <0.003 | | 9970 | 8327–11,614 |
| G + A | 1182.9 | 3.5 | 1 | 0.061 | | 10,213 | 8416–12,009 |
| G + A + N | 1131.7 | 61.1 | 5 | <0.001 | | 16,129 | 9973–22,286 |
| G + A + N + R | 1133.0 | 0.7 | 1 | 0.403 | | 16,188 | 9983–22,394 |

$^*P$-value for likelihood-ratio test.

We now turn to the results for the Zelterman regression model, presented in the bottom panel of Table 7. Here the model with Gender, Age and Nation is also the best model in terms of model fit as well as AIC. If we compare the population size estimates under the truncated Poisson regression model with those under the Zelterman regression model, we find that, for models with identical covariates, the population size estimates under the Zelterman model are much larger. This suggests that the Zelterman model corrects for the downward bias in the population size estimates from the truncated Poisson regression model when overdispersion is present.

## APPENDIX: VARIANCE ESTIMATION UNDER COVARIATES

We now provide details for computing a variance estimate of the generalized Zelterman estimator (3.4), which we write as

$$\hat{N}_Z = \sum_{i=1}^n \frac{1}{w_i} = \sum_{i=1}^N \frac{\Delta_i}{w_i},$$

where $w_i = 1 - \exp(-2e^{\hat{\eta}_i})$ and $\Delta_i$ is an indicator which is 1 (unit is sampled) with probability $w_i$ and 0 (unit is not sampled) with probability $1 - w_i$.

We use the techniques of *conditioning* to develop a variance estimator of (3.4) and follow the methodological development in van der Heijden et al. (2003a). We have that [see Ross (1985), page 125]

$$(A.1) \qquad \mathrm{Var}(\hat{N}_Z) = \mathrm{Var}_n[E(\hat{N}_Z|n)] + E_n[\mathrm{Var}(\hat{N}_Z|n)],$$

where moments inside the brackets are computed *conditional* upon $n$ and moments outside the bracket refer to the marginal distribution of $n$. Consider $E(\hat{N}_Z|n)$ and its estimate

$$\widehat{E(\hat{N}_Z|n)} = \sum_{i=1}^n \frac{1}{w_i} = \sum_{i=1}^N \frac{\Delta_i}{w_i}.$$

Consequently,

$$\mathrm{Var}_n\left(\sum_{i=1}^N \frac{\Delta_i}{w_i}\right) = \sum_{i=1}^N \mathrm{Var}_n\left(\frac{\Delta_i}{w_i}\right) = \sum_{i=1}^N w_i(1-w_i)/w_i^2 = \sum_{i=1}^N (1-w_i)/w_i,$$

for which an unbiased estimator can be provided as

$$(A.2) \qquad \widehat{\mathrm{Var}}_n\left(\sum_{i=1}^N \frac{\Delta_i}{w_i}\right) = \sum_{i=1}^N \Delta_i(1-w_i)/w_i^2 = \sum_{i=1}^n (1-w_i)/w_i^2.$$

We move on to consider the second term, $E_n[\mathrm{Var}(\hat{N}_Z|n)]$, involved in (A.1). We write

$$(A.3) \qquad \mathrm{Var}(\hat{N}_Z|n) = \mathrm{Var}\left(\sum_{i=1}^N \frac{\Delta_i}{w_i}\Big|\Delta_1,\ldots,\Delta_N\right),$$

so that

$$\mathrm{Var}(\hat{N}_{\mathrm{Z}}|n) = \mathrm{Var}\left(\sum_{i=1}^{n} \frac{1}{w_i}\right).$$

Recall that $w_i = 1 - \exp(v_i)$ and $v_i = -2e^{\hat{\eta}_i}$, so that

$$w_i = w_i(\hat{\beta}) = 1 - \exp(v_i) = 1 - \exp(-2e^{\hat{\eta}_i}) = 1 - \exp(-2e^{\hat{\beta}^T \mathbf{x}_i}).$$

Consequently, $w_i(\hat{\beta})$ and $w_j(\hat{\beta})$ will *not* be independent for $i \neq j$, since both depend on a common $\hat{\beta}$. An application of the multivariate $\delta$-method as done by van der Heijden et al. (2003a) provides

$$(\mathrm{A}.4) \qquad \left(\sum_i \nabla w_i(\hat{\beta})^T\right) Cov(\hat{\beta}) \left(\sum_i \nabla w_i(\hat{\beta})\right),$$

where

$$(\mathrm{A}.5) \qquad \nabla w_i(\hat{\beta}) = \frac{(1 - w_i)v_i}{w_i^2} \mathbf{x}_i.$$

Summing (3.5) and (A.4) give the full variance approximation of $\mathrm{Var}(\hat{N}_{\mathrm{Z}})$.

**Acknowledgments.** We would like to express our deepest thanks to the Editor, the Associate Editor and two reviewers for their helpful comments and suggestions. We are also grateful to James Gallagher (Statistical Services Center, University of Reading) for providing many helpful comments on the final version of the paper.

## SUPPLEMENTARY MATERIAL

**Computer programmes and illegal immigrant data** (DOI: 10.1214/08-AOAS214SUPP; .zip).

APPLIED STATISTICS
SCHOOL OF BIOLOGICAL SCIENCES
UNIVERSITY OF READING
PHILIP LYLE BUILDING
WHITEKNIGHTS
READING RG6 6BX
UNITED KINGDOM
E-MAIL: d.a.w.bohning@reading.ac.uk

DEPARTMENT OF METHODOLOGY
  AND STATISTICS
FACULTY OF SOCIAL AND
  BEHAVIOURAL SCIENCES
UTRECHT UNIVERSITY
P.O. BOX 80.140
3508 TC UTRECHT
THE NETHERLANDS
E-MAIL: p.g.m.vanderheijden@uu.nl